\documentclass[a4paper,aps,prl,showpacs,twocolumn]{revtex4-1}      
\usepackage[utf8]{inputenc}  
\usepackage[T1]{fontenc}     
\usepackage[british]{babel}  
\usepackage[sc,osf]{mathpazo}
\usepackage[scaled=0.86]{berasans}  
\usepackage[scaled=1.03]{inconsolata} 
\usepackage[usenames,dvipsnames]{color} 
\usepackage[colorlinks,citecolor=magenta,linkcolor=green,urlcolor=blue]{hyperref}  
\usepackage{graphicx} 
\usepackage[babel]{microtype}  
\usepackage{amsmath,amssymb,amsthm,bm,mathtools,amsfonts,mathrsfs,bbm} 
\usepackage{xspace}  
\usepackage{pgfplots}  
\DeclareMathOperator{\tr}{tr}
\DeclareMathOperator{\Tr}{Tr}

\newcommand{\bra}[1]{\left\langle #1 \right|}
\newcommand{\ket}[1]{\left| #1 \right\rangle}

\newcommand{\ba}{\begin{eqnarray}}
\newcommand{\ea}{\end{eqnarray}}
\newcommand{\ban}{\begin{eqnarray*}}
\newcommand{\ean}{\end{eqnarray*}}

\begin{document}

\title{Incompatible quantum measurements admitting a local hidden variable model}

\author{Marco T\'ulio Quintino}
\affiliation{D\'epartement de Physique Th\'eorique, Universit\'e de Gen\`eve, 1211 Gen\`eve, Switzerland}
\author{Joseph Bowles}
\affiliation{D\'epartement de Physique Th\'eorique, Universit\'e de Gen\`eve, 1211 Gen\`eve, Switzerland}
\author{Flavien Hirsch}
\affiliation{D\'epartement de Physique Th\'eorique, Universit\'e de Gen\`eve, 1211 Gen\`eve, Switzerland}
\author{Nicolas Brunner}
\affiliation{D\'epartement de Physique Th\'eorique, Universit\'e de Gen\`eve, 1211 Gen\`eve, Switzerland}

\date{\today}  

\begin{abstract}
The observation of quantum nonlocality, i.e. quantum correlations violating a Bell inequality, implies the use of incompatible local quantum measurements. Here we consider the converse question. That is, can any set of incompatible measurements be used in order to demonstrate Bell inequality violation? Our main result is to construct a local hidden variable model for an incompatible set of qubit measurements. Specifically, we show that if Alice uses this set of measurements, then for any possible shared entangled state, and any possible dichotomic measurements performed by Bob, the resulting statistics are local. This represents significant progress towards proving that measurement incompatibility does not imply Bell nonlocality in general.
\end{abstract}

\maketitle

A key aspect of quantum theory is that certain observables cannot be jointly measured, in strong contrast with classical physics. This leads to many prominent quantum features, such as the uncertainty principle and information gain vs disturbance trade-off, and plays a central role in quantum information processing \cite{chuang}. The incompatibility of quantum observables is usually captured via the notion of commutativity: incompatible observables do not commute. However, quantum theory allows for more general measurements, so-called positive-operator valued measures (POVM), the incompatibility of which cannot be properly captured using commutativity \cite{krausbook}. Here a natural concept is that of joint measurability \cite{buschbook}. A set of POVMs is said to be jointly measurable if each one of them can be derived from coarse-graining of one common POVM. Conversely, if such a joint POVM does not exist, the set is considered incompatible. The concept of joint measurability thus arguably provides a natural separation between classical and non-classical sets of measurements. 

A long-standing question is to understand the relation between the incompatibility of quantum measurements and quantum nonlocality \cite{bell64,brunnerreview}, another key feature of quantum theory. When performing a set of well-chosen measurements on a shared entangled state, two distant observers can observe nonlocal correlations, i.e. which cannot be explained by a local (i.e. classical) model. The question is then how the non-classicality of quantum measurements (i.e. their incompatibility) relates to the non-classicallity of quantum correlations, detected via violation of a Bell inequality. While the observation of nonlocality implies the use of incompatible measurements (for both observers), the converse is not known. Specifically, the question is the following. For any possible set of incompatible measurements performed by one observer, can we always find a shared entangled state and a set of measurements for the second observer, such that the resulting statistics will lead to Bell inequality violation?

\begin{figure}[b!]
\includegraphics[width=0.9\columnwidth]{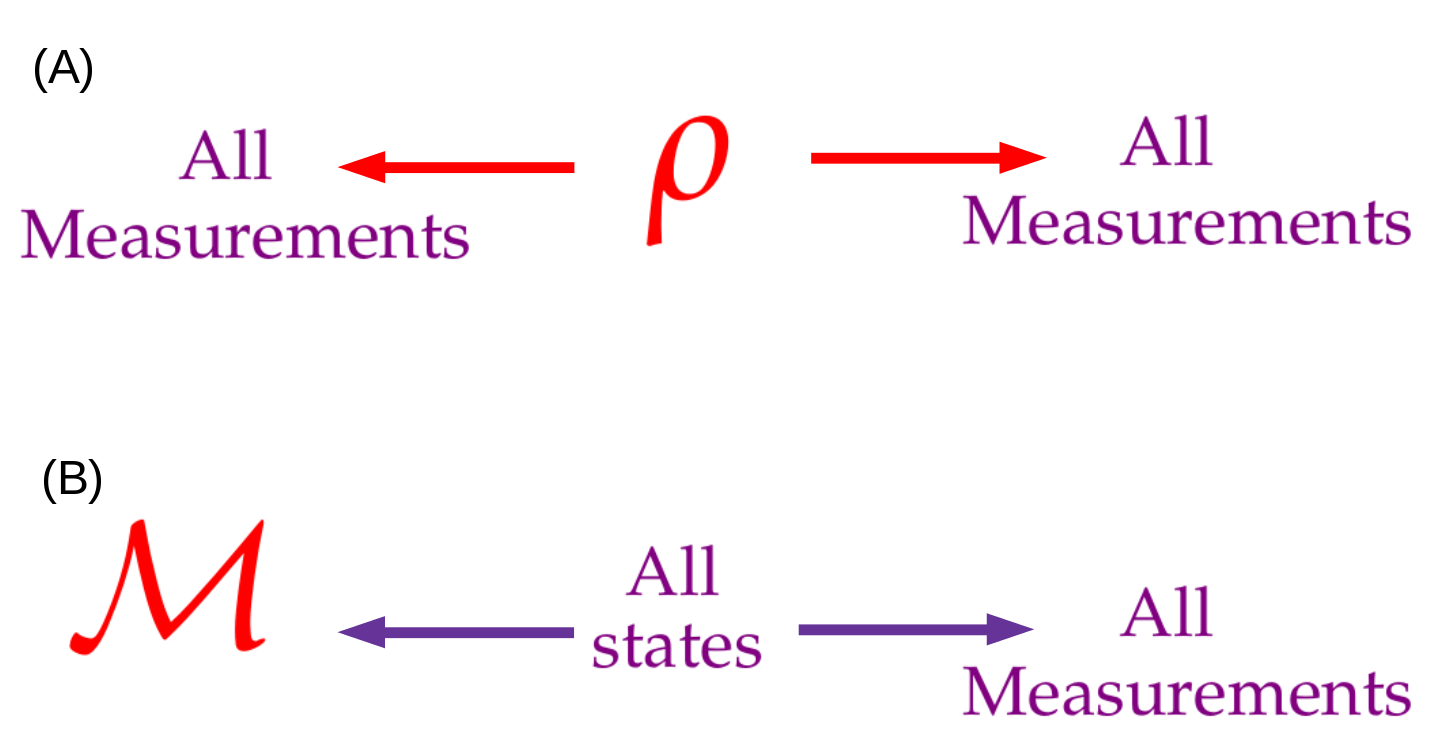}
\caption{\label{fig0} The problem of classically simulating quantum correlations has two facets. (A) Constructing a LHV model for a given entangled quantum state $\rho$, considering arbitrary local measurements for Alice and Bob. (B) Constructing a LHV model for a given set of incompatible measurements $\mathcal{M}$ (performed by Alice), considering arbitrary entangled states, and arbitrary local measurements Bob. While question (A) has been extensively studied, much less is known about question (B), which is the focus of this work.}
\end{figure}

In the case of projective measurements, the answer is positive as proven many years ago \cite{tsirelson85}. 
For the case of POVMs, however, the question is much more difficult. In the simplest case of two dichotomic POVMs, Wolf et al. \cite{wolf09} proved that incompatibility is equivalent to violation of the Clauser-Horne-Shimony-Holt \cite{chsh69} inequality, confirming previous evidence \cite{anderson05,son05}. However, their proof cannot be extended to the general case, as the joint measurability problem cannot be reduced to a pair of POVMs only \cite{krausbook}. For instance, it is possible to have a set of three POVMs which is incompatible, although any pair (among the three) is jointly measurable \cite{teiko08,liang11}. Recently, a strong connection between joint measurability and EPR steering \cite{wiseman07}, a form of quantum nonlocality strictly weaker than Bell nonlocality \cite{quintino15}, has been demonstrated \cite{uola14,quintino14,uola15}, leading to interesting results in both areas \cite{teiko15}. More generally, the connection between measurement uncertainty and nonlocality in no-signaling theories has been discussed \cite{oppenheim10,pfister13,banik15}.

In the present work we show that a set of incompatible quantum measurements can admit a LHV model. Specifically, we consider a bipartite Bell test in which Alice performs a given non-jointly measurable set of qubit POVMs. We then show that the statistics of such an experiment, considering an arbitrary shared entangled state and any possible dichotomic measurements performed by Bob, can be exactly reproduced using only classical shared resources. In other words this set of incompatible measurements, despite having some nonclassical feature, can never lead to nonlocal correlations (considering dichotomic measurements for Bob). A parallel can be drawn to the study, initiated by Werner \cite{werner89}, of quantum states which are entangled (hence non-classical), but nevertheless admit a LHV model; see e.g. \cite{barrett02,acin06,almeida07,hirsch13,finite_SR,joe} and \cite{augusiak_review} for a recent review. In contrast, we show that a set of nonclassical measurements admits a LHV model (see Fig.1). Finally, we discuss the perspective of extending our result to the most general Bell test, which would thus demonstrate that incompatibility does not imply Bell nonlocality in general.

\textit{Preliminaries.} 
We start by introducing concepts and notations. Consider a set of $N$ POVMs, given by operators $M_{a|x}$ satisfying $ \sum_a M_{a|x} = \openone$, $M_{a|x}\geq 0$ for $x \in \{1, \ldots , N \}$. This set is said to be jointly measurable if there exists one common POVM, $M_{\vec{a}}$, with outcomes $\vec{a}=[a_{x=1},a_{x=2},\ldots,a_{x=N}]$ where $a_{x}$ gives the outcome of measurement $x$, that is
\begin{align}
         M_{\vec{a}} \geq  0 , \quad \sum_{\vec{a}} M_{\vec{a}} =  \openone , \quad
        \sum_{\vec{a}\setminus a_x}M_{\vec{a}} = M_{a\vert x}  \;,
\end{align}
where $\vec{a}\setminus a_x$ stands for the elements of $\vec{a}$ except for $a_x$. Hence, all POVM elements $M_{a|x}$ are recovered as marginals of the \textit{joint observable} $M_{\vec{a}}$. Notably, joint measurability of a set of POVMs does not imply that they commute \cite{kru}. Moreover, partial joint measurability does not imply full joint measurability in general \cite{krausbook}, contrary to commutation. More generally, any partial compatibility configuration can be realized in quantum theory \cite{fritz14}.

The focus of this work is to connect the incompatibility of a set of measurement to quantum nonlocality. We thus consider a Bell scenario featuring two observers, Alice and Bob, sharing an entangled state $\rho$. Alice and Bob perform local measurements, represented by operators $M_{a|x}$ and $M_{b|y}$. Here $x$ and $y$ denote the choice of measurement settings, while $a$ and $b$ denote the outcomes. The resulting probability distribution is thus given by $p(ab|xy) = \tr(\rho M_{a|x} \otimes M_{b|y})$. This distribution is local (in the sense of Bell) if it admits a decomposition of the form
\begin{equation} \label{LHV}
    p(ab \vert xy)= \int d \lambda q(\lambda) p_A(a\vert x,\lambda) p_B(b \vert y,\lambda).
\end{equation}
Here the local model consists of a classical (hidden) variable $\lambda$, distributed according to density $q(\lambda)$, and Alice's and Bob's local response functions represented by the probability distributions $p_A(a\vert x,\lambda) $ and $ p_B(b \vert y,\lambda)$. On the contrary, if a decomposition of the form \eqref{LHV} cannot be found, the distribution $p(ab|xy)$ is termed nonlocal, and violates (at least) one Bell inequality \cite{bell64,brunnerreview}.

It is straightforward to show that if the set of Alice's measurements, $\mathcal{M}_A = \{M_{a|x} \}$, is jointly measurable, the resulting distribution $p(ab|xy) $ is local, for any possible entangled state $\rho$ and arbitrary measurements of Bob; see e.g. \cite{quintino14}. Indeed, if the set $\mathcal{M}_A$ is compatible, then Alice can recover all statistics from one joint observable. Clearly, no Bell inequality violation can be obtained if Alice always performs the same measurement. 

The main goal of this work is to discuss the converse problem. Specifically, given that the set $\mathcal{M}_A $ is incompatible, what can we say about the locality of the distribution $p(ab \vert xy)$? Previous work \cite{wolf09} demonstrated a striking connection in the simplest case, when $\mathcal{M}_A $ consists of two dichotomic POVMs. Any set $\mathcal{M}_A $ that is not jointly measurable can be used to demonstrate nonlocality. Whether this connection holds for more general sets of POVMs has been an open question since then. Here we show that, for certain incompatible sets of POVMs, the resulting distribution $p(ab \vert xy)$ is always local, considering arbitrary entangled states $\rho$ and arbitrary dichotomic measurements on Bob's side \cite{footnote1}.

\textit{Main result.} 
We consider the continuous set of dichotomic qubit POVMs, $\mathcal{M}_A^\eta  = \{ M_{\pm| \hat{x}}^\eta\}$, with elements
\ba \label{POVM}
M_{\pm| \vec{x}}^\eta = \frac{1}{2}  (  \openone  \pm \eta \,  \hat{x} \cdot \vec{\sigma})
\ea
with binary outcome $a=\pm1$. Here $\hat{x} $ is any vector on the Bloch sphere denoting the measurement direction, and $\vec{\sigma} = (\sigma_1, \sigma_2, \sigma_3)$ is the vector of Pauli matrices. Note that the set $\mathcal{M}_A^\eta$ features a parameter $0 \leq \eta \leq 1$, representing basically the purity of the POVM elements. For $\eta=1$, all POVM elements are projectors 
\ba 
\Pi_{\pm| \hat{x}} = \frac{1}{2} (  \openone   \pm \,  \hat{x} \cdot \vec{\sigma}).
\ea 
The set $\mathcal{M}_A^{\eta=1}$ is simply the set of all qubit projective measurements, and is thus clearly incompatible. For $\eta=0$, the set contains only the identity (thus clearly compatible). In general the set $\mathcal{M}_A^\eta$ contains noisy measurements, with elements simply given by $M_{\pm| \hat{x}}^\eta = \eta \Pi_{\pm| \hat{x}} + (1-\eta) \openone/2 $. In fact, the set $\mathcal{M}_A^\eta$ is jointly measurable if and only if $\eta\leq 1/2$ \cite{quintino14,uola14}. 

Below we will show that there is $\eta^*> 1/2$ such that the set $\mathcal{M}_A^{\eta^*}$ is local in any Bell test, considering arbitrary states $\rho$ and arbitrary dichotomic measurements for Bob. Since $\mathcal{M}_A^{\eta^*}$ is not jointly measurable, this shows that incompatibility is not sufficient for Bell inequality violation in this case. Below we give a full proof of the result, proceeding in several steps.

The first step consists in exploiting the symmetries of the problem in order to find the minimal set of states $\rho$ we need to consider. By linearity of the problem---the probabilities $p(ab|xy)$ are linear in $\rho$, and the set of local correlations is convex, see e.g. \cite{brunnerreview}---we can safely focus on pure states. Indeed, if there was a mixed state $\rho$ leading to Bell inequality violation using measurements $\mathcal{M}_A^{\eta^*}$, there would also be a pure state doing so. 

Next, given that $\mathcal{M}_A^{\eta^*}$ consists only of qubit measurements, Alice's subsystem can be considered to be a qubit.
Moreover, since we are free to choose convenient local reference frames (i.e. we can apply any local unitaries on Alice and Bob's systems), the shared state $\rho$ (of dimension $2 \times d$) can therefore be expressed in the Schmidt form \cite{chuang}, i.e. $\rho = \ket{\phi_\theta} \bra{\phi_\theta}$ with
\ba \label{pure_ent}
\ket{\phi_\theta} = \cos{\theta} \ket{00} +  \sin{\theta} \ket{11}
\ea
and $\theta \in [ 0 , \pi/4]$.

Now we introduce the measurements on Bob's side. Since Bob's system is of rank 2, we can focus here on dichotomic qubit measurements. As any such POVM can be viewed as a projective qubit measurement followed by classical post-processing \cite{dariano05}, it is sufficient to discuss projective qubit measurements $\Pi_{b|\hat{y}} =  (  \openone  + b  \,  \hat{y} \cdot \vec{\sigma})/2$, where $\hat{y}$ is any vector on the Bloch sphere and $b=\pm1$. 

Our goal is thus to show that there exists $\eta^*>1/2$ such that the distribution 
\ba 
p(ab|xy) = \tr( \ket{\phi_\theta} \bra{\phi_\theta} M_{a|\hat{x}}^{\eta^*} \otimes \Pi_{b|\hat{y}})
\ea
is local for any measurement directions $\hat{x}$ and $\hat{y}$, and any state $\ket{\phi_\theta}$. In other words we would like to construct a LHV model for the incompatible set of measurements $\mathcal{M}_A^{\eta^*}$. In order to do so, we start by reformulating the problem by making use of the following relation:
\ba \label{equiv}
\tr( \ket{\phi_\theta} \bra{\phi_\theta} M_{a|\hat{x}}^{\eta} \otimes \Pi_{b|\hat{y}}) = 
\tr( \rho_\theta^{\eta} \Pi_{a|\hat{x}} \otimes \Pi_{b|\hat{y}})
\ea
where 
\ba \label{rhoMaf}
\rho_\theta^{\eta} = \eta \ket{\phi_\theta} \bra{\phi_\theta} + (1- \eta ) \frac{\openone}{2} \otimes \rho_B
\ea
and $\rho_B = \tr_A(\ket{\phi_\theta} \bra{\phi_\theta} )$. Thus, the problem of constructing a LHV model for $\mathcal{M}_A^{\eta^*}$ (considering dichotomic measurements for Bob) is equivalent to the problem of constructing a LHV model for the class of states $\rho_\theta^{\eta^*}$ (for all $\theta \in [ 0 , \pi/4]$) with arbitrary projective measurements for Alice and Bob. Importantly, it must be shown that 
$\rho_\theta^{\eta^*}$ admit a LHV model for all $\theta \in [ 0 , \pi/4]$ and for a fixed $\eta^*>1/2$ (independent of $\theta$).

The locality of the states $\rho_\theta^{\eta^*}$ must be discussed in two steps, for different ranges of the parameter $\theta $. First consider the range $\theta \in [ 0 , \pi/4- \epsilon] $ with $\epsilon>0$. Recently, we presented a sufficient condition for a two-qubit state to admit a LHV model for projective measurements \cite{joe}. For states of the form $\rho_\theta^{\eta}$, a LHV model was shown to exist given that  
\ba \label{condition}
\cos^2(2\theta)  \geq \frac{2\eta-1}{(2-\eta)\eta^3}.
\ea 
Hence for any $\theta$, we get a corresponding value of $\eta$ for which the state is provably local; see Fig.2. This clearly guarantees that for $\theta \in [ 0 , \pi/4- \epsilon] $, with $\epsilon>0$ fixed, we can find $\eta^*>1/2$ such that $\rho_\theta^{\eta^*}$ is local. However, when $\theta$ gets closer to $\pi/4$, this approach will not work. Indeed, there is no fixed value $\eta^*>1/2$ for which locality can be guaranteed for any $\theta \in [ 0 , \pi/4]$, as can be seen by continuity of Eq. \eqref{condition} or from Fig.2. We thus need to find a different approach for this regime.

We proceed as follows. First note that for the case $\theta = \pi/4$, the state $\rho_\theta^{\eta}$ is simply a two-qubit Werner state 
\ba
\rho_W^\mu = \mu \ket{\phi_{+}} \bra{\phi_+} + (1- \mu ) \frac{\openone}{4} 
\ea with $\ket{\phi_+} = (\ket{00}+ \ket{11})/\sqrt{2}$. Coincidently such states admit a LHV model for $\mu \leq \mu_{LHV} \simeq 0.66$, considering arbitrary projective measurements \cite{acin06}. The case $\theta=\pi/4$ is thus covered. Let us next discuss the case of $\theta$ in the neighborhood of $\pi/4$. To do so we consider the problem of decomposing the target state $\rho_\theta^{\eta}$ as a mixture of states admitting a LHV model. Specifically, we demand for which values of $\theta$ and $\eta$, we can find a convex combination of the form:
\ba \label{problem}
\rho_\theta^{\eta} = \alpha \rho_W^{\mu_{LHV}} + (1-\alpha) \sigma
\ea
with $0 \leq \alpha \leq 1$. Here $\sigma$ is an unspecified two-qubit state, which we are free to choose. As long as $\sigma$ admits a LHV model, this implies that $ \rho_\theta^{\eta}$ is local. In order to do so, we simply ensure that 
\ba 
\sigma =  \frac{\rho_\theta^{\eta} - \alpha \rho_W^{\mu_{LHV}}}{1-\alpha} 
\ea
is a valid separable state. By setting $\alpha = \frac{1}{\mu_{LHV}} \eta  \sin(2 \theta)$, we obtain a diagonal matrix $\sigma$ (for all $\eta$ and $\theta$). It is straightforward to check that the eigenvalues of $\sigma$ are positive when 
\ba 
\eta \leq \frac{\mu_{LHV} }{(1+\mu_{LHV}) \cot{\theta}- \mu_{LHV} }.
\ea
By combining condition \eqref{condition} and the above result, it follows that the state $\rho_\theta^{\eta}$ admits a LHV model for any $\theta$ and for $\eta \leq \eta^* \simeq 0.503$. Note that a better bound can be obtained using numerical methods. Consider again the problem of finding a decomposition of the form \eqref{problem} with $\sigma$ a separable state. For fixed $\theta$, the optimal decomposition can be found via semi-definite programming (SDP): 
\begin{align} \label{SDP}
& \text{max } \eta \\
\text{s.t. }&\rho_\theta^{\eta} = \alpha \rho_W^{\mu_{LHV}} +  \sigma \nonumber\\ \nonumber
&\sigma\geq0 , \quad \sigma^{PT} \geq 0  , \quad  \Tr{\sigma}+ \alpha = 1, \quad \alpha \geq 0.
\end{align}
Here $\sigma^{PT}$ denotes the partial transpose \cite{peres} of $\sigma$. Verifying that $\sigma^{PT}$ is positive ensures here that $\sigma$ is separable \cite{horodecki_ppt}. The result of this optimization procedure is shown on Fig.2. Combining again with condition \eqref{condition} we get that $\rho_\theta^{\eta}$ admits a LHV model for $\eta \leq \eta^* \simeq 0.515$ (for any $\theta$), for all projective measurements for Alice and Bob.

\begin{figure}
\includegraphics[width=\columnwidth]{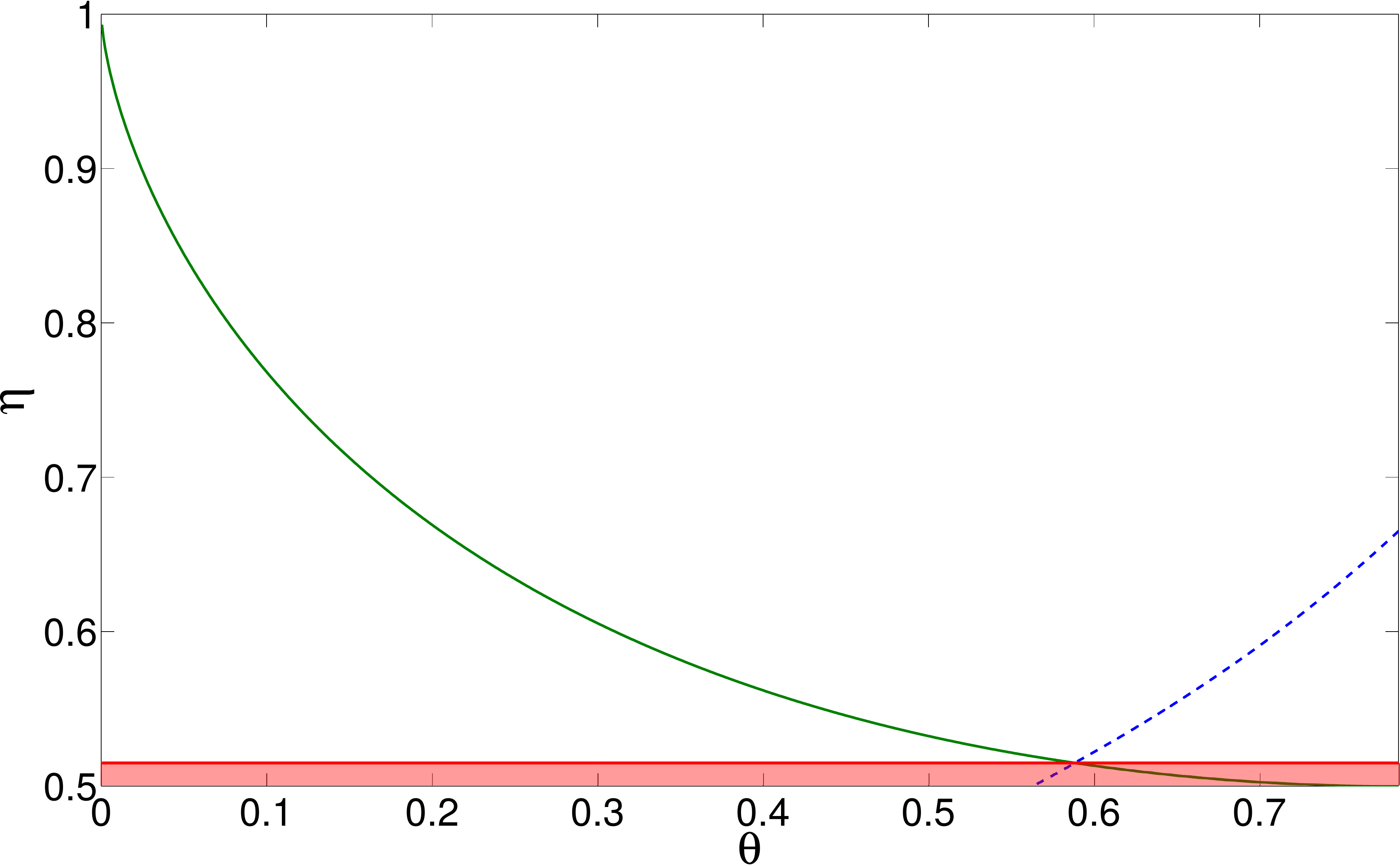}
\caption{\label{fig1} Parameter region for which the state $\rho_\theta^{\eta}$ admits a LHV model. First, below the green curve, as given by Eq. \eqref{condition}. Second below the blue dashed curve, as found via the SDP \eqref{SDP}. The two curves cross at $\eta^* \simeq 0.515$. It follows that the state $\rho_\theta^{\eta}$ is local for $ \eta \leq \eta^*$ and for all $\theta$, i.e. in the shaded region, below the red horizontal line. }
\end{figure}

We therefore conclude that in the range $1/2 < \eta^* \lesssim 0.515$, the set of measurements $\mathcal{M}_A^{\eta^*}$ is incompatible and admits a LHV model. Specifically, $\mathcal{M}_A^{\eta^*}$ can never lead to Bell inequality violation, considering arbitrary shared entangled states and arbitrary dichotomic measurements performed by the second observers. 

Finally, it is worth mentioning that this result can be straightforwardly extended to the case of a set containing only a finite number of incompatible measurements. For instance, we have checked that a set of 12 well-chosen POVMs in $\mathcal{M}_A^{\eta}$ (chosen rather uniformly on the Bloch sphere) is incompatible for $\eta > 0.512$, via standard SDP techniques \cite{wolf09}. However this set clearly admits a LHV model for $\eta  \lesssim 0.515$. 

It would be interesting to see if the result also holds in the simplest case of a set of only 3 POVMs. Consider for instance the 3 Pauli operators: $\sigma_x$, $\sigma_y$, and $\sigma_z$. Adding noise as in Eq. \eqref{POVM}, the resulting POVMs are pairwise jointly measurable, but still not fully jointly measurable, in the range $1/\sqrt{3} <  \eta \leq 1/\sqrt{2}$ \cite{teiko08,liang11}. Could such a set of 3 POVMs admit a LHV model?

\textit{Discussion.}  We discussed the relation between measurement incompatibility and Bell nonlocality. Specifically, we showed that a given set of incompatible qubit measurements can never lead to Bell inequality violation, as it admits a LHV model. Our construction covers the case of any possible shared entangled state, and all possible dichotomic measurements performed by the second observer. 

The main open question now is whether our result can be extended to non-dichotomic measurements on Bob's side. If possible, this would then prove that measurement incompatibility does not imply Bell nonlocality in general \cite{footnote2}. 

We believe that the prospects for extending our LHV model for the set of measurements $\mathcal{M}_A^{\eta}$ to general measurements on Bob's side is promising. More precisely, following our approach, this amounts to show that the states $\rho_\theta^{\eta}$ of Eq. \eqref{rhoMaf} (for a fixed $\eta>1/2$ and all $\theta$) admit a LHV model, considering arbitrary projective measurements for Alice and arbitrary POVMs for Bob \cite{footnote3}.
We conjecture that this is the case, which is also supported by the fact that, so far, there is no example of an entangled state admitting a LHV model for projective measurements, but not for POVMs.

\emph{Acknowledgements.} We thank Matt Pusey and Tam\'as V\'ertesi for discussions. We acknowledge financial support from the Swiss National Science Foundation (grant PP00P2\_138917 and Starting grant DIAQ).



\begin{thebibliography}{10}

\bibitem{chuang}
M.~Nielsen and I.~Chuang, {\em Quantum Computation and Quantum Information}.
\newblock Cambridge Series on Information and the Natural Sciences. Cambridge
  University Press, 2000.

\bibitem{krausbook}
K.~Kraus, A.~Bohm, J.~Dollard, and W.~Wootters, {\em States, effects, and
  operations: fundamental notions of quantum theory}.
\newblock Lecture notes in physics. Springer-Verlag, 1983.

\bibitem{buschbook}
P.~Busch, P.~Lahti, and P.~Mittelstaedt, {\em The Quantum Theory of
  Measurement}.
\newblock No.~v. 2 in Environmental Engineering. Springer, 1996.

\bibitem{bell64}
J.~S. Bell, ``{On the Einstein-Poldolsky-Rosen paradox},'' {\em Physics}
  {\bfseries 1}, 195--200 (1964).

\bibitem{brunnerreview}
N.~{Brunner}, D.~{Cavalcanti}, S.~{Pironio}, V.~{Scarani}, and S.~{Wehner},
  ``{Bell nonlocality},''
  \href{http://dx.doi.org/10.1103/RevModPhys.86.419}{{\em Reviews of Modern
  Physics} {\bfseries 86}, 419--478 (2014)}.

\bibitem{tsirelson85}
L.~A. Khalfin and B.~S. Tsirelson, ``Quantum and quasi-classical analogs of
  Bell inequalities,'' {\em Symposium on the Foundations of Modern Physics}
  441--460 (1985).

\bibitem{chsh69}
J.~F. Clauser, M.~A. Horne, A.~Shimony, and R.~A. Holt, ``Proposed Experiment
  to Test Local Hidden-Variable Theories,''
  \href{http://dx.doi.org/10.1103/PhysRevLett.23.880}{{\em Phys. Rev. Lett.}
  {\bfseries 23}, 880--884 (1969)}.

\bibitem{wolf09}
M.~M. {Wolf}, D.~{Perez-Garcia}, and C.~{Fernandez}, ``{Measurements
  Incompatible in Quantum Theory Cannot Be Measured Jointly in Any Other
  No-Signaling Theory},''
  \href{http://dx.doi.org/10.1103/PhysRevLett.103.230402}{{\em Physical Review
  Letters} {\bfseries 103}, 230402 (2009)}.

\bibitem{anderson05}
E.~{Andersson}, S.~M. {Barnett}, and A.~{Aspect}, ``{Joint measurements of
  spin, operational locality, and uncertainty},''
  \href{http://dx.doi.org/10.1103/PhysRevA.72.042104}{{\em Phys. Rev.~A}
  {\bfseries 72}, 042104 (2005)}.

\bibitem{son05}
W.~{Son}, E.~{Andersson}, S.~M. {Barnett}, and M.~S. {Kim}, ``{Joint
  measurements and Bell inequalities},''
  \href{http://dx.doi.org/10.1103/PhysRevA.72.052116}{{\em Phys. Rev.~A}
  {\bfseries 72}, 052116 (2005)}.

\bibitem{teiko08}
T.~{Heinosaari}, D.~{Reitzner}, and P.~{Stano}, ``{Notes on Joint Measurability
  of Quantum Observables},''
  \href{http://dx.doi.org/10.1007/s10701-008-9256-7}{{\em Foundations of
  Physics} {\bfseries 38}, 1133--1147 (2008)}.

\bibitem{liang11}
Y.-C. {Liang}, R.~W. {Spekkens}, and H.~M. {Wiseman}, ``{Specker's parable of
  the overprotective seer: A road to contextuality, nonlocality and
  complementarity},''
  \href{http://dx.doi.org/10.1016/j.physrep.2011.05.001}{{\em Phys. Rep.}
  {\bfseries 506}, 1--39 (2011)}.

\bibitem{wiseman07}
H.~M. {Wiseman}, S.~J. {Jones}, and A.~C. {Doherty}, ``{Steering, Entanglement,
  Nonlocality, and the Einstein-Podolsky-Rosen Paradox},''
  \href{http://dx.doi.org/10.1103/PhysRevLett.98.140402}{{\em Phys. Rev. Lett.}
  {\bfseries 98}, 140402 (2007)}.

\bibitem{quintino15}
M.~T. Quintino, T.~V\'ertesi, D.~Cavalcanti, R.~Augusiak, M.~Demianowicz,
  A.~Ac\'{\i}n, and N.~Brunner, ``Inequivalence of entanglement, steering, and
  Bell nonlocality for general measurements,''
  \href{http://dx.doi.org/10.1103/PhysRevA.92.032107}{{\em Phys. Rev. A}
  {\bfseries 92}, 032107 (2015)}.
 

\bibitem{uola14}
R.~{Uola}, T.~{Moroder}, and O.~{G{\"u}hne}, ``{Joint Measurability of
  Generalized Measurements Implies Classicality},''
  \href{http://dx.doi.org/10.1103/PhysRevLett.113.160403}{{\em Phys. Rev.
  Lett.} {\bfseries 113}, 160403 (2014)}.

\bibitem{quintino14}
M.~T. {Quintino}, T.~{V{\'e}rtesi}, and N.~{Brunner}, ``{Joint Measurability,
  Einstein-Podolsky-Rosen Steering, and Bell Nonlocality},''
  \href{http://dx.doi.org/10.1103/PhysRevLett.113.160402}{{\em Phys. Rev.
  Lett.} {\bfseries 113}, 160402 (2014)}.

\bibitem{uola15}
R.~{Uola}, C.~{Budroni}, O.~{G{\"u}hne}, and J.-P. {Pellonp{\"a}{\"a}}, ``{A
  one-to-one mapping between steering and joint measurability problems},'' 
  {{\em Phys. Rev. Lett.} {\bfseries 115}, 230402 (2015)}.

\bibitem{teiko15}
T.~{Heinosaari}, J.~{Kiukas}, D.~{Reitzner}, and J.~{Schultz},
  ``{Incompatibility breaking quantum channels},''  J. Phys. A: Math. Theor. {\bf 48}, 435301 (2015).

\bibitem{oppenheim10}
J.~{Oppenheim} and S.~{Wehner}, ``{The Uncertainty Principle Determines the
  Nonlocality of Quantum Mechanics},''
  \href{http://dx.doi.org/10.1126/science.1192065}{{\em Science} {\bfseries
  330}, 1072-- (2010)}.

\bibitem{pfister13}
C.~{Pfister} and S.~{Wehner}, ``{An information-theoretic principle implies
  that any discrete physical theory is classical},''
  \href{http://dx.doi.org/10.1038/ncomms2821}{{\em Nature Communications}
  {\bfseries 4}, 1851 (2013)}.

\bibitem{banik15}
M.~{Banik}, ``{Measurement incompatibility and
  Schr{\"o}dinger-Einstein-Podolsky-Rosen steering in a class of probabilistic
  theories},'' \href{http://dx.doi.org/10.1063/1.4919546}{{\em Journal of
  Mathematical Physics} {\bfseries 56}, 052101 (2015)}.

\bibitem{werner89}
R.~F. Werner, ``Quantum states with Einstein-Podolsky-Rosen correlations
  admitting a hidden-variable model,''
  \href{http://dx.doi.org/10.1103/PhysRevA.40.4277}{{\em Phys. Rev. A}
  {\bfseries 40}, 4277--4281 (1989)}.

\bibitem{barrett02}
J.~{Barrett}, ``{Nonsequential positive-operator-valued measurements on
  entangled mixed states do not always violate a Bell inequality},''
  \href{http://dx.doi.org/10.1103/PhysRevA.65.042302}{{\em Phys. Rev.~A}
  {\bfseries 65}, 042302 (2002)}.

\bibitem{acin06}
A.~{Ac{\'{\i}}n}, N.~{Gisin}, and B.~{Toner}, ``{Grothendieck's constant and
  local models for noisy entangled quantum states},''
  \href{http://dx.doi.org/10.1103/PhysRevA.73.062105}{{\em Phys. Rev.~A}
  {\bfseries 73}, 062105 (2006)}.

\bibitem{almeida07}
M.~L. {Almeida}, S.~{Pironio}, J.~{Barrett}, G.~{T{\'o}th}, and
  A.~{Ac{\'{\i}}n}, ``{Noise Robustness of the Nonlocality of Entangled Quantum
  States},'' \href{http://dx.doi.org/10.1103/PhysRevLett.99.040403}{{\em Phys.
  Rev. Lett.} {\bfseries 99}, 040403 (2007)}.

\bibitem{hirsch13}
F.~{Hirsch}, M.~T. {Quintino}, J.~{Bowles}, and N.~{Brunner}, ``{Genuine Hidden
  Quantum Nonlocality},''
  \href{http://dx.doi.org/10.1103/PhysRevLett.111.160402}{{\em Physical Review
  Letters} {\bfseries 111}, 160402 (2013)}.

\bibitem{finite_SR}
J.~{Bowles}, F.~{Hirsch}, M.~T. {Quintino}, and N.~{Brunner}, ``{Local Hidden
  Variable Models for Entangled Quantum States Using Finite Shared
  Randomness},'' \href{http://dx.doi.org/10.1103/PhysRevLett.114.120401}{{\em
  Physical Review Letters} {\bfseries 114}, 120401 (2015)}.

\bibitem{joe}
J.~{Bowles}, F.~{Hirsch}, M.~T. {Quintino}, and N.~{Brunner}, ``{Sufficient
  criterion for guaranteeing that a two-qubit state is unsteerable},'' Phys. Rev. A {\bf 93}, 022121 (2016).
  
\bibitem{augusiak_review}
R.~{Augusiak}, M.~{Demianowicz}, and A.~{Ac{\'{\i}}n}, ``{Local
  hidden--variable models for entangled quantum states},''  J. Phys. A {\bf 42}, 424002 (2014).

\bibitem{kru}
P.~Kruszyski and W.~de~Muynck, ``Compatibility of observables represented by
  positive operator-valued measures,'' {\em Journal of mathematical physics}
  {\bfseries 28}, 1761--1763 (1987).

\bibitem{fritz14}
R.~{Kunjwal}, C.~{Heunen}, and T.~{Fritz}, ``{All joint measurability
  structures are quantum realizable},'' Phys. Rev. A {\bf 89}, 052126 (2014).
  
  \bibitem{footnote1}
  One could also consider a simpler question, where both Alice and Bob use the same set of measurements. 
  In this case, however, the problem is much easier, and examples of incompatible sets not leading to Bell violation can be found in Ref. \cite{quintino14}.

\bibitem{dariano05}
G.~{Mauro D'Ariano}, P.~{Lo Presti}, and P.~{Perinotti}, ``{Classical
  randomness in quantum measurements},''
  \href{http://dx.doi.org/10.1088/0305-4470/38/26/010}{{\em Journal of Physics
  A Mathematical General} {\bfseries 38}, 5979--5991 (2005)}.


\bibitem{peres}
A.~{Peres}, ``{Separability Criterion for Density Matrices},''
  \href{http://dx.doi.org/10.1103/PhysRevLett.77.1413}{{\em Phys. Rev.
  Lett.} {\bfseries 77}, 1413--1415 (1996)}.

\bibitem{horodecki_ppt}
M.~{Horodecki}, P.~{Horodecki}, and R.~{Horodecki}, ``{Separability of mixed
  states: necessary and sufficient conditions},''
  \href{http://dx.doi.org/10.1016/S0375-9601(96)00706-2}{{\em Phys. Lett.
  A} {\bfseries 223}, 1--8 (1996)}.

\bibitem{footnote2} One may also wonder about multipartite Bell tests. Consider for instance a tripartite Bell test, where A performs measurements $\mathcal{M}_A^{\eta}$. Clearly, nonlocality can be obtained if B and C simply ignore A, and make appropriate measurements on a shared singlet. Hence the relevant question here is whether nonlocality can be obtained on the partition $A|BC$, which brings us back to a bipartite Bell test (with arbitrary outputs on Bob's side).

\bibitem{footnote3} Note that it is sufficient to consider the cases of three and four output qubit POVMs, as extremal qubit POVMs have (at most) four outcomes \cite{dariano05}.


\end{thebibliography}

\providecommand{\href}[2]{#2}\begingroup\raggedright\endgroup

\end{document}